\documentclass[paper]{JHEP3}

\usepackage{epsfig,bbm}

\newcommand{\be}{\begin{equation}} 
\newcommand{\ee}{\end{equation}}
\newcommand{\bea}{\begin{eqnarray}} 
\newcommand{\eea}{\end{eqnarray}}

\newcommand{\bmp}{\noindent\begin{minipage}{16cm}}
\newcommand{\emp}{\end{minipage}\vskip 7mm} 
\def\lsim{\mathrel{\raise.3ex\hbox{$<$\kern-.75em\lower1ex\hbox{$\sim$}}}}
\def\gsim{\mathrel{\raise.3ex\hbox{$>$\kern-.75em\lower1ex\hbox{$\sim$}}}}

\newcommand{\intron}[1]{}

\title{The Next Generation}

\author{Oleg Antipin\footnote{oleg.a.antipin@jyu.fi} and Matti Heikinheimo\footnote{matti.p.s.heikinheimo@jyu.fi}\\
        \\ Department of Physics, P.O.Box 35 (YFL), 
        \\ FI-40014 University of Jyv\"askyl\"a, Finland, 
        \\ and 
  	    \\ Helsinki Institute of Physics, P.O.~Box 64, 
  	    \\ FI-00014 University of Helsinki, Finland.\\}
\author{Kimmo Tuominen \footnote{kimmo.tuominen@jyu.fi}\,\,\footnote{On leave of absence from Department of physics, University of Jyv\"askyl\"a} \\
{CP}$^{ \bf 3}${-Origins}, 
Campusvej 55,\\ DK-5230 Odense M, Denmark \\and
Helsinki Institute of Physics, 
P.O.Box 64, \\ FI-000140, University of Helsinki, Finland}

\abstract{We consider the possibility of non-sequential generation(s) of Standard Model  -like matter as a consequence of cancellation of global and gauge anomalies due to a new strongly interacting sector responsible for the electroweak symmetry breaking. We consider concrete models for the strong dynamics and outline several scenarios for the next generation. For these we provide analysis of the precision constraints as well as a discussion on collider signatures and implications for cosmology.}


\preprint{CP3-Origins-2010-04}

\begin{document}

\section{Introduction}
\label{intro}
With the commissioning of the Large Hadron Collider (LHC), the possible experimental discovery of physics beyond the Standard Model (SM) is becoming reality. Gathering and analysis of large amount of data hopefully finally sheds light towards the correct model building paradigms as well as properties and spectra of new beyond SM (BSM) matter fields. One appealing model building paradigm has been the dynamical breaking of electroweak symmetry, i.e. Technicolor (TC) \cite{TC,Hill:2002ap}; for up-to-date reviews see \cite{Sannino:2008ha,Sannino:2009za}.

Phenomenologically viable Technicolor theories are those featuring so-called walking dynamics: as a function of the energy scale, the Technicolor gauge coupling evolves very slowly, walks, over a large hierarchy of scales. Recently a new class of walking technicolor theories was introduced in \cite{Sannino:2004qp}. These theories are minimal in the sense that walking is achieved with very small amount of new matter fields by utilizing higher fermion representations. It was also shown that these theories are compatible with precision electroweak constraints and the (composite) Higgs particle can be light \cite{Dietrich:2005jn}; see also \cite{Doff:2009na,Doff:2009kq}. Phenomenology studies of these models have been carried out both for collider signatures \cite{Belyaev:2008yj, Foadi:2007se,Foadi:2007ue, Antipin:2009ch,Frandsen:2009fs,Antipin:2009ks} and dark matter \cite{Foadi:2008qv,Gudnason:2006ug,Gudnason:2006yj,Kainulainen:2006wq,Kouvaris:2007ay,Khlopov:2007ic,Khlopov:2008ty,Kouvaris:2008hc,Kainulainen:2009rb}.

The basic matter content underlying walking dynamics, as proposed in \cite{Sannino:2004qp} and which we will also consider in this paper, consists of SU($N$) gauge dynamics with matter in the two-index symmetric representation. We will denote theory with $N$ technicolors and $N_f$ techniflavors in the two-index symmetric representation by TC($N$,$N_f$-2S). In particular we will concentrate on two- and three-color theories TC(2,2-2S) and TC(3,2-2S). When the technifermions in TC(2,2-2S) are assigned electroweak quantum numbers, the number of weak doublets is odd resulting in an ill-defined theory \cite{Witten:1982fp}. As noted in \cite{Sannino:2004qp} this anomaly can be simply cured by adding one doublet, singlet under technicolor and ordinary QCD color. Hence, from the weak interaction viewpoint the techniquarks and this new doublet of leptons form a natural fourth family of matter. Gauge anomaly cancellation constrains the hypercharge assignments within this new fourth generation, but not uniquely\cite{Sannino:2004qp,Dietrich:2005jn}: For example one can consider hypercharges under which the techniquarks carry electric charges of the usual QCD quarks and the fourth generation leptons are a heavy replica of an ordinary lepton generation. But more exotic possibilities exist, like doubly or even fractionally charged leptons. The consequences of sequential fourth generation have been investigated in the
literature in the past, see \cite{Frampton:1999xi} for a review. Recently there has
been renewed interest into the subject, see e.g. \cite{Holdom:2006mr,Kribs:2007nz,Belotsky:2002ym,Belotsky:2004ga,Belotsky:2008se}. However, as the examples mentioned above show, there is motivation for much richer variety of alternatives. If new QCD quark or SM-like lepton generations exist beyond the observed three, they can appear non-sequential. As in the example discussed above, there may be a fourth generation of leptons but no corresponding QCD quarks.

Motivated by this simple example, in this paper we exhibit several other possibilities and discuss the associated phenomenology in detail. We will consider concretely the underlying Technicolor dynamics to be given by TC(2,2-2S) and TC(3,2-2S). Then, we add SM-like matter (i.e. new QCD quark or SM-like lepton generations) in order to cancel gauge anomalies and maintain the overall number of weak doublets even. We show that this leads to numerous natural and novel possibilities. In addition to the already investigated case where there is a fourth generation of leptons but no QCD quarks, these include the case where there is a fourth generation of QCD quarks but no new leptons and the more exotic possibilities where more than one new generation of quarks and/or leptons appear. Out of the various possibilities we pick the most interesting scenarios for which we consider the phenomenological implications.

\section {The Next Generation from Walking Technicolor models}
\label{models}

For the technicolor sector we will consider concrete models TC(2,2-2S)  (two flavors in the adjoint of SU(2)) and TC(3,2-2S) (two flavors in the sextet of SU(3)) to illustrate various possibilities for the next generation. In these theories the TC matter charged under SM is arranged into electroweak doublets in the standard way. We will consider the TC matter to have SM-like hypercharges (i.e. mimicking either leptons or quarks) and then find the extra (TC singlet) SM-like matter content needed to cancel Witten and gauge anomalies. We will list several possibilities and discuss the physics related to the most interesting candidate cases in more detail.

To fix the notation, in the TC(2,2-2S) and TC(3,2-2S) models we consider, the technicolored matter fields constitute one electroweak doublet
\be
Q^a_L=\left(\begin{array}{c} U^a_L\\ D^a_L\end{array}\right)
\ee
with $a=1,\dots,d(R)$ where $d(R)$ is the dimension of the representation of the technicolor gauge group to which the fermions have been assigned. The right handed fields $U^a_R$ and $D^a_R$ are singlets under the weak SU$_L(2)$. Now, following the logic specified above, we will explore different hypercharge assignments $Y(Q_L)$ with the hypercharges of the right-handed fields fixed by $Y(U_R)=Y(Q_L)+T_3(U_L)$ and $Y(D_R)=Y(Q_L)+T_3(D_L)$. Then, for a given assignment of the hypercharge of the techniquarks, we compensate for the gauge anomalies by adding further techni-singlet matter fields with quantum numbers of either SM-like leptons or QCD quarks. Since the SU$_L(2)$ gauge theory suffers from the Witten anomaly, in addition to canceling gauge anomalies we must maintain the number of electroweak doublets even. 

The anomaly with two external TC gauge bosons and one external U$_Y(1)$ boson vanishes as does also the gravitational anomaly. The nontrivial conditions to study are the one with two SU$_L(2)$ and one U$_Y(1)$ gauge boson, denoted by $A(2,2,1)$, and the one with three external U$_Y(1)$ gauge bosons, denoted by $A(1,1,1)$. Hence, we need to solve
\bea
A(2,2,1) &=& -\frac{1}{2}(N_D(l)-N_D(q))+d_{\rm{TC}}Y(Q_L)=0 \\
A(1,1,1) &=& -\frac{3}{4}(N_D(q)-N_D(l))+d_{\rm{TC}}(2Y^3(Q_L)-Y^3(U_R)-Y^3(D_R))=0.
\eea
 Here $N_D(l)$ and $N_D(q)$ denote, respectively, the number of SM-like leptons and QCD quarks required to be introduced in order to cancel the gauge anomalies due to the presence of technifermions. The dimension of the technifermion representation is denoted by $d_{\rm{TC}}$. It turns out that these equations imply only a constraint on the difference $\Delta N\equiv N_D(l)-N_D(q)$. For generic $Y(Q_L)$ the solution is
 \be
 \Delta N=2d_{\rm{TC}} Y(Q_L).
 \ee
 To make further progress, we must pick $Y(Q_L)$. The above relation already provides some restrictions since $\Delta N$ must be an integer and our choice of the underlying TC model corresponds to $d_{\rm{TC}}=3,6$ for TC(2,2-2S) and TC(3,2-2S) respectively. We choose to investigate the values which are already present in the SM, namely we consider $Y(Q_L)=\pm 1/2,\pm 1/6$. Of course this may still seem rather {\em ad hoc}, but on the other hand fermions are natural and the models of this type are logical extensions of Technicolor theories. Our motivation for the particular choice of the above mentioned hypercharges is on one hand the fact that these hypercharge assignments have been already considered in studies of TC(2,2-2S). On the other hand some of these values are interesting in terms of global symmetries; The SM is known to enjoy a discrete $Z_6$ symmetry which can be enlarged to TC sector only for particular values of the techniquark hypercharge \cite{Zubkov:2007yw}. For TC(2,2-2S) this can be done for $Y(Q_L)=\pm 1/2$ and for TC(3,2-2S) for $Y=-1/6$. 
  
Then, for these hypercharge assignments we have $\Delta N=\pm 1, \pm 3$ for TC(2,2-2S) (i.e. $d_{\rm{TC}}=3$) and  $\Delta N=\pm 2, \pm 6$ for TC(3,2-2S) ($d_{\rm{TC}}=6$) corresponding to $Y(Q_L)=\pm 1/6, \pm 1/2$, respectively, in both cases. Phenomenologically the most interesting are those where only few new doublets are introduced. This is due to constraints from precision data through oblique corrections. We will now single out some interesting special cases and then study those in more detail.

Let us start with the TC(2,2-2S) model for which the most interesting possibilities seem to be
\begin{itemize}
\item $Y(Q_L)=1/2$ whence we introduce three new SM-like lepton families.
\item $Y(Q_L)=-1/2$ whence we introduce three new QCD-quark families. This will result in large additional contribution to (S,T) parameters, and possibly put QCD close to an infrared fixed point. We will not consider this possibility further. 
 \item $Y(Q_L)=1/6$ whence we introduce one SM-like lepton family ($Y(L_L)=-1/2$). This is the scenario considered in \cite{Antipin:2009ks}.
 \item $Y(Q_L)=-1/6$ whence we introduce one SM quark family ($Y(q_L)=1/6$). In this scenario {\it the definition} for the new left-handed quark doublet is $q_L\equiv (\bar{d},\bar{u})^T$. This amounts to rewrite the SU(2)$_L$ theory for our new quarks in the c.c. representation of SU(2)$_L$. In this new formulation,  right-handed antibottom of the next generation will play a role of a left-handed top in the usual formulation.
\end{itemize}
Of these the third and fourth are most interesting. Note that in all these cases when taking into account the technifermions and the new SM-like matter fields, the overall number of electroweak doublets is even and hence there is no Witten anomaly. The details of the third one have been investigated elsewhere  \cite{Antipin:2009ks} and the fourth one will be investigated in the next section.

Similarly we consider TC(3,2-2S) where we consider most interesting the following cases:

\begin{itemize}
\item $Y(Q_L)=1/2$ whence we need to introduce six SM-like lepton families. We will not consider this further.
\item $Y(Q_L)=-1/2$ whence we need to introduce six QCD quark families. We will not consider this further.
\item $Y(Q_L)=1/6$ whence we need to introduce two SM-like lepton families. 
\item $Y(Q_L)=-1/6$ whence we need to introduce two QCD quark families. 
\end{itemize}

Of these, the third one is very interesting and will be studied in more detail in the next section. Note again that the overall number of weak doublets is even.

Finally, we point out that in the above analysis we have considered adding matter only with SM-like quantum numbers. This assumption can be of course relaxed. For example for TC(2,2-2S) and hypercharge $Y(Q_L)=\pm 1/2$ we may introduce doubly charged lepton family, i.e. the doublet $L_L$ and the singlets $E_R$ and $N_R$ with $Y(L_L)=\mp 3/2$ to cancel the anomalies. We will consider the phenomenology of this case briefly also in the next section. Of course one can consider different realization for the dynamical symmetry breaking model than the TC(2,2-2S) and TC(3,2-2S).

\section{Constraints and Collider Signals}
\label{constraints}

As already discussed in the previous sections, if next generation of SM-like matter fields appears in LHC, it may be very different than a simple extrapolation from the three known generations. We have presented a model framework where either one or more lepton families may arise but no QCD quarks or instead a new QCD quark generation may appear but no associated fourth generation leptons. Let us now discuss the constraints on the masses of these new matter fields and physical consequences of their existence.

\subsection{TC(2,2-2S) and New QCD quarks}
\label{QCD_quarks}
A new QCD quark generation would be interesting: in addition to possible direct observation at LHC it may help to explain some recent observations on CP-phenomena at B-factories and it will significantly enhance the Higgs production cross section at LHC via the gluon fusion channel.

The basic constraint in this type of models is obtained from the precision electroweak data, i.e. oblique corrections. Let us first discuss the oblique corrections resulting from adding a new QCD-quark family with $Y(q_L)=1/6$. For Dirac particles, the contributions to $S$, $T$ and $U$ parameters can be obtained analytically, and the required formulae can be found in \cite{He:2001tp}. For the mass degenerate constituent techniquarks, the contribution to $S$ is perturbatively estimated to be

\begin{equation}
S_{TC}=\frac{N_f/2}{6\pi}d(R),
\end{equation}
where $N_f$ is the number of techniquarks and $d(R)$ is the dimension of the representation of the techniquarks. The walking behavior is expected to reduce this value via nonperturbative corrections \cite{Appelquist:1998xf}. In the following analysis, we take this reduction to be $30$\%.

\begin{figure}
\includegraphics[width=0.49\textwidth]{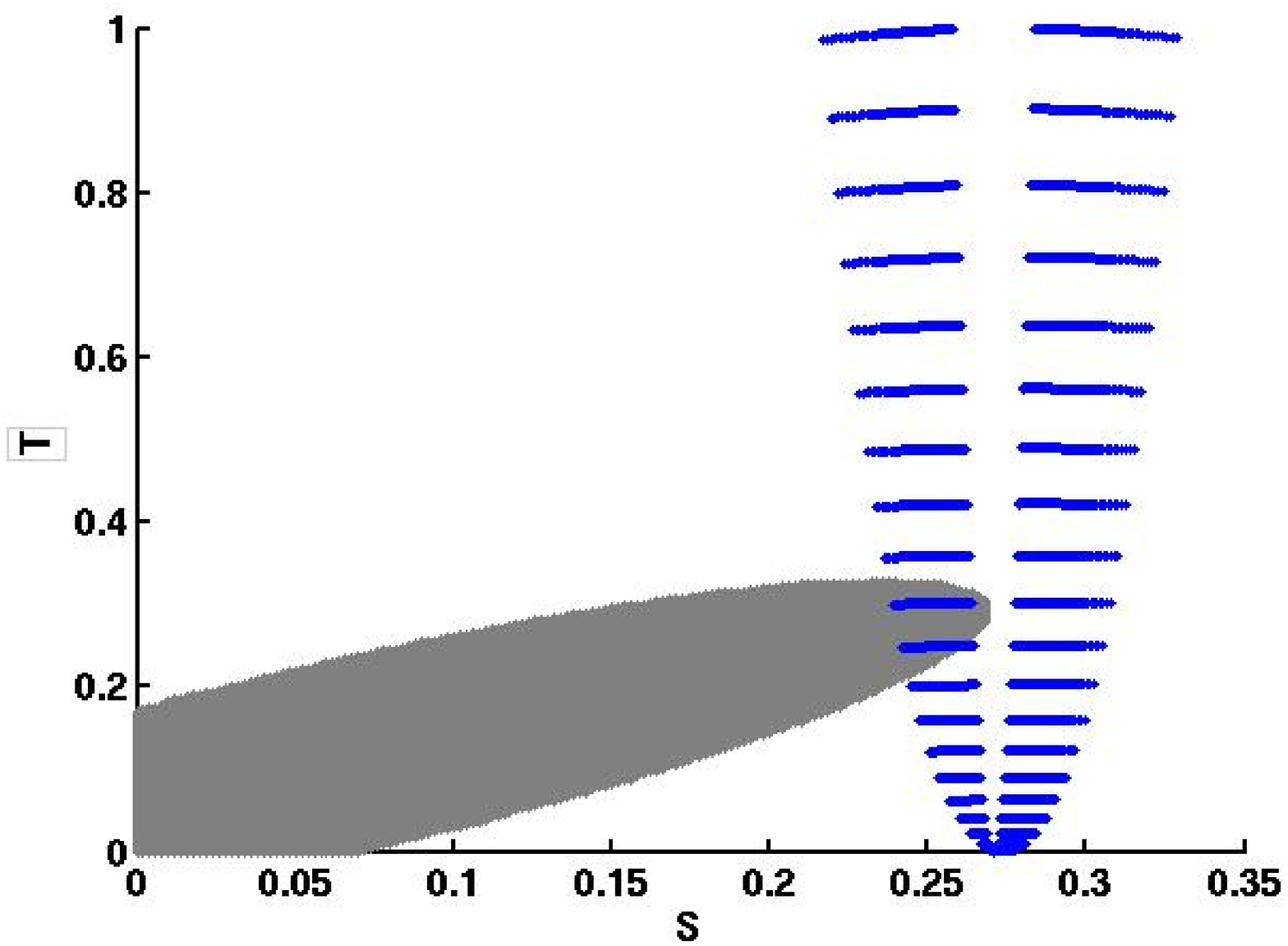}
\includegraphics[width=0.49\textwidth]{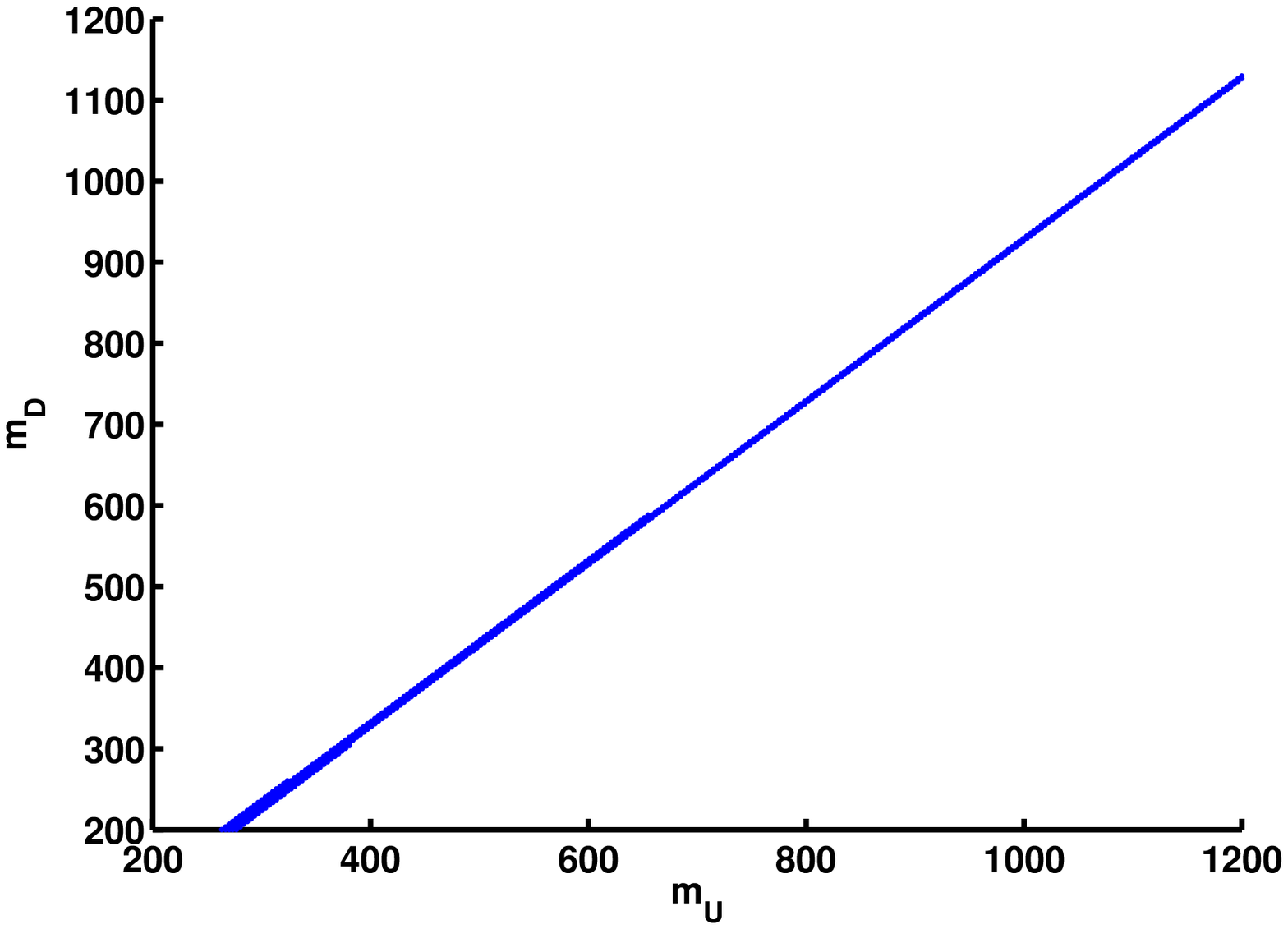}
\caption{Left panel: The $(S,T)$-spectrum of a new QCD-quark family, and the allowed $3\sigma$-region. Right panel: The allowed masses of the quarks.}
\label{quarks}
\end{figure}

The resulting values of $S$ and $T$ are shown in the left panel of Figure \ref{quarks}. The gray ellipsis shows the $3\sigma$-allowed region obtained from the LEPEWWG \cite{:2005ema}. The right panel shows the allowed masses for the new quarks, with $S$ and $T$ inside the ellipse and $U<0.05$; due to the small value of $U$ the comparison with LEPEWWG data where $U=0$ is enforced is justified. We see that the quarks are allowed to have a wide range of masses, but the mass difference is limited by the $T$-parameter to the range $m_u-m_d \sim 50-75$ GeV.

The existence of the fourth QCD quark generation would enhance the (composite) Higgs effective coupling to two gluons. Here we imagine that from the low energy viewpoint the composite Higgs arising from the underlying technicolor model would act as an effective SM-like complex scalar doublet. For the case of T(2,2-2S) this has been shown to be the case due to the hypercharge conservation even if the scalar sector has a more complex structure \cite{Foadi:2007ue}, while for the T(3,2-2S) this is expected directly on the basis of the scalar spectrum and the global flavor symmetry breaking pattern.

\begin{figure}
\begin{center}
\includegraphics[width=0.49\textwidth]{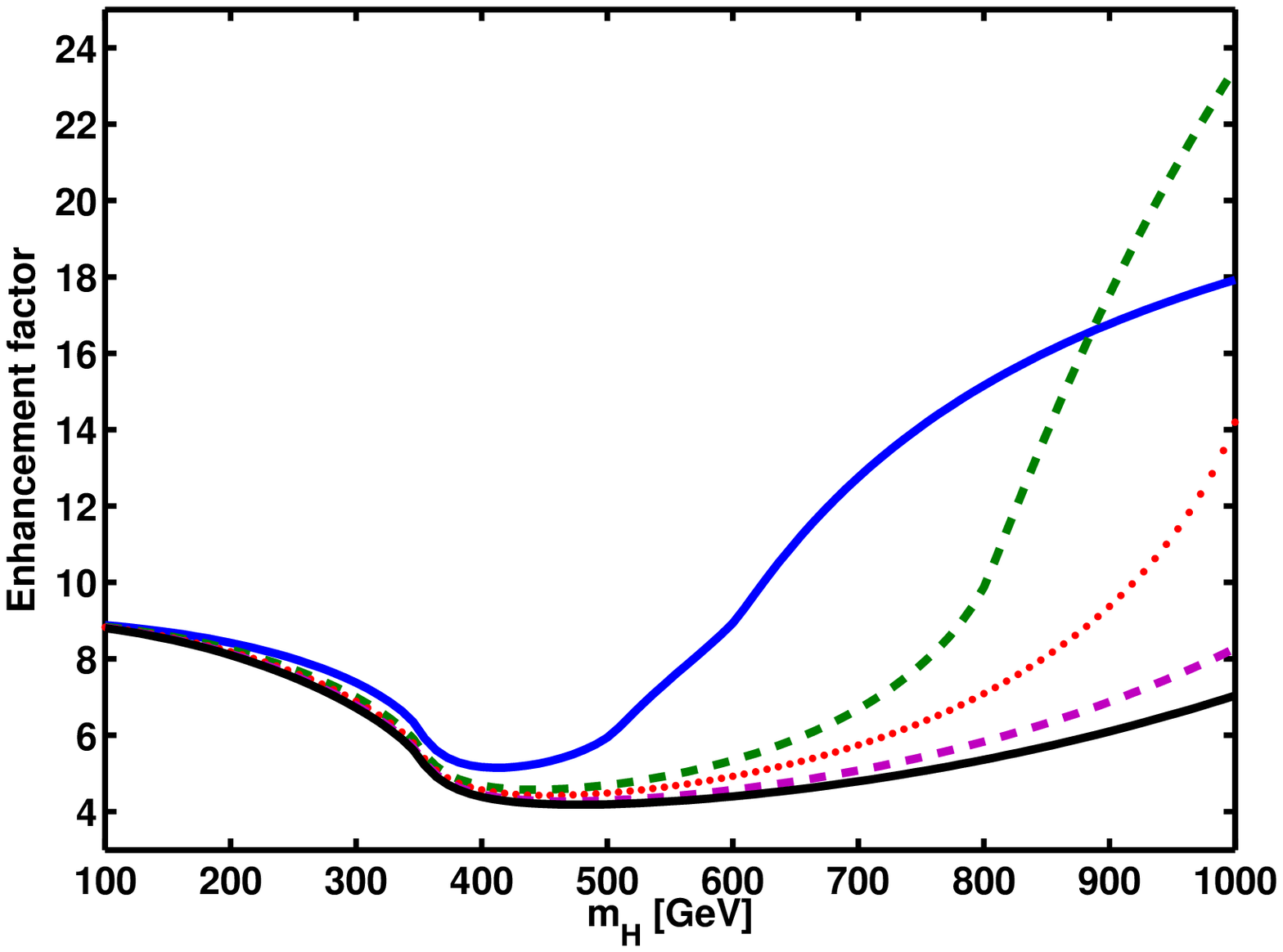}
\includegraphics[width=0.49\textwidth]{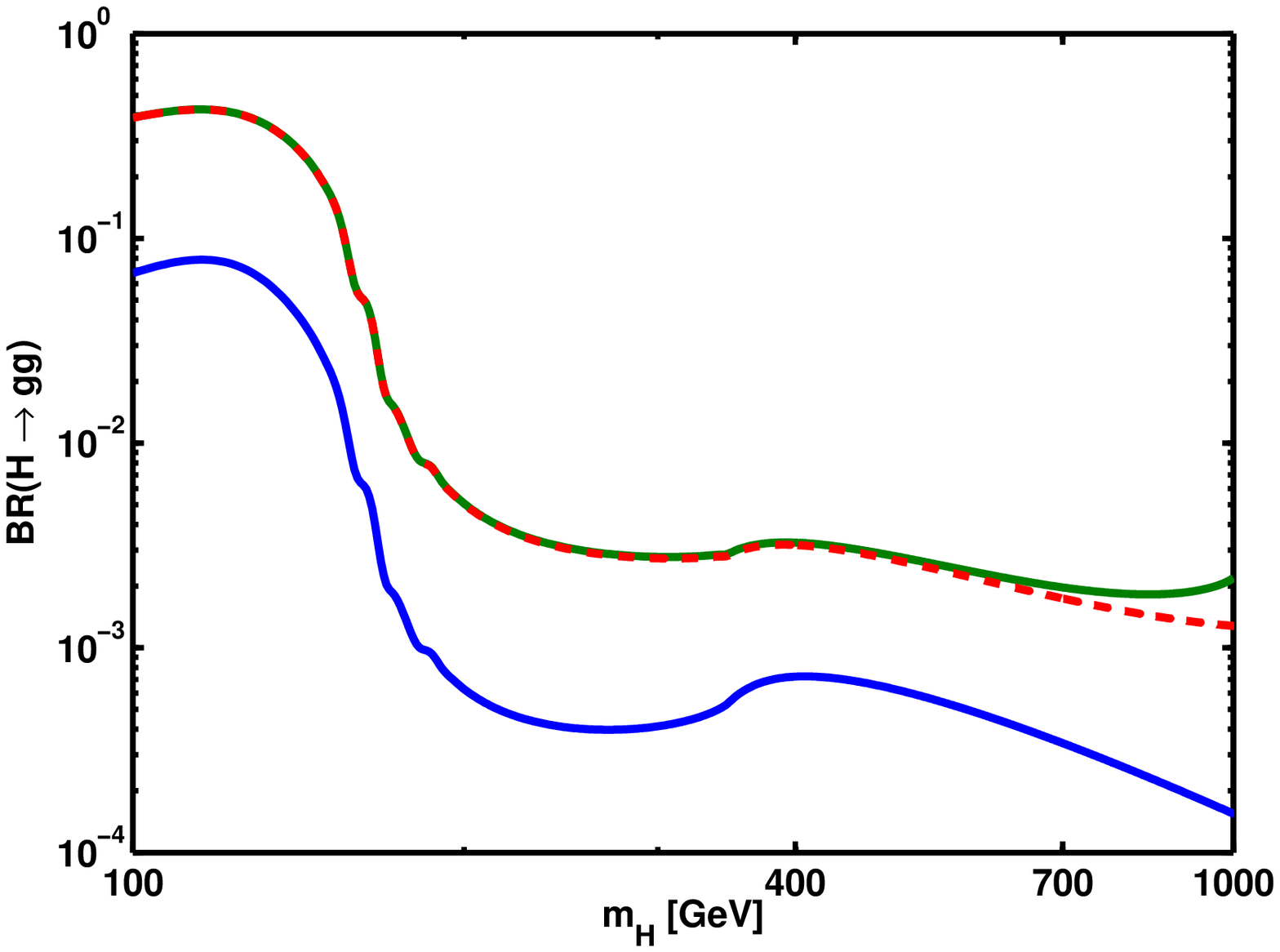}
\caption{Left panel: The enhancement factor $\sigma_4(gg\rightarrow H)/\sigma_{SM}(gg\rightarrow H)$ as a function of the Higgs mass for quark masses $(m_d,m_u)=(250,300)$ GeV (blue solid line), (400,400) GeV (green dashed line), (500,500) GeV (red dotted line), (800,850) GeV (purple dashed line) and $(\infty ,\infty )$ (black solid line). Right panel: The Higgs decay branching fraction $\Gamma(H\rightarrow gg)/\Gamma(H\rightarrow X)$ with fourth generation quark masses (500,500) GeV (green solid line) and (800,850) GeV (red dashed line), and in the SM (blue solid line).}
\label{Higgsprod}
\end{center}
\end{figure}

The presence of the new heavy quarks will affect both the Higgs production via the gluon fusion channel and the Higgs decay to two gluons. This coupling is generated by a heavy quark loop with two gluon legs and one Higgs leg. In the SM, the only significant contribution to the $Hgg$-coupling comes from the top quark loop. The presence of two new heavy quarks would thus enhance the coupling by approximately a factor of three, yielding a nine times as large cross section as in the SM. This effect has been extensively studied in the literature, and the relevant cross section formulae can be found for example in \cite{Dawson:1990zj}. The actual value of the enhancement factor depends on the masses of the new quarks and on the mass of the Higgs particle. The behavior of the enhancement factor as a function of the Higgs mass for a few given values of the quark masses is shown in the left panel of Figure \ref{Higgsprod}. The right panel shows the effect of the new quarks to Higgs decay branching fraction into the two gluon channel. The implications of a new QCD quark generation for Higgs searches in the Tevatron and at the LHC have been studied in \cite{Schmidt:2009kk}.

The fourth QCD quark generation, if within the accessible mass range, will provide interesting signals at the LHC. On the basis of the constraints from the precision data, let us assume that $m_U>m_D$. Due to the small mass splitting between $U$ and $D$ the decay $U\rightarrow DW^\ast$ can only occur into a virtual $W$-boson. Then, depending on the mass of the $D$ quark, it would decay to $t$ accompanied either with a real or virtual $W$-boson. In addition to the charged current decay, one can also have a flavor-changing neutral current decay $D\rightarrow b + Z$ at one loop \cite{Barger:1986ha,Hou:1988sx}. The role of this process is crucially dependent on the mass of the new $D$ quark, which is constrained by present searches for fourth generation to be $m_D\ge 200$ GeV \cite{Amsler:2008zzb}.  As soon as the mass of the $D$ quark exceeds $\sim 250$ GeV, the two body decay into $tW$ becomes accessible and immediately also becomes the dominant mode. Hence there is a very little window for the $bZ$ mode and probably the most interesting early warning process would be $pp \to U \bar{U} \to D \bar{D} + 2 W \to b \bar{b} + 6 W$ leading to the signal involving the jets, two same-sign leptons and missing energy \cite{Holdom:2006mr,Holdom:2007nw}. 

In addition to the possible direct observation at the LHC, the presence of the fourth QCD quark generation may already have appeared indirectly in connection with the CP-violation phenomena \cite{Soni:2008bc}: Recall that generally for $N_g$ quark generations the Cabibbo-Kobayashi-Maskawa (CKM) mixing matrix is parametrized in terms of $N_g(N_g-1)/2$ angles and $(N_g-1)(N_g-2)/2$ phases when the freedom to redefine the quark fields has been taken into account to eliminate $2N_g-1$ real parameters from the  total number $N_g^2$ of real parameters required for the parametrization of $N_g\times N_g$ unitary matrix. In contrast to the single phase responsible for all CP-violating phenomena in the case of three quark generations, there are three phases in the case of four generations. The two extra phases imply possible extra sources for the CP-violation \cite{Andriyash:2005ax}, and in \cite{Soni:2008bc} it was shown that this extra freedom indeed provides adequate explanation for the recent indications of new physics involved in CP asymmetries of the $b$-quark as observed in the B-factories. Moreover, the existence of this additional CP-violation may play a role in establishing the origin of matter-antimatter asymmetry through the electroweak baryogenesis. Constraints from the existing data on the properties of the fourth generation quarks suggest that the mass of the fourth generation top, i.e. the new quark with electric charge +2/3, has mass within the range 400-600 GeV \cite{Soni:2008bc,Soni:2010xh}, which fits in nicely with the constraints from the oblique corrections we have obtained here.

Another possible scenario, as noted in the previous section, is to introduce two new generations of QCD quarks with TC(3,2-2S). This would produce $S\gtrsim 0.45$ in the region where $T$ is small, and is thus most likely ruled out. Accompanied with some new physics that produces a large negative $T$, like Majorana neutrinos (which also give negative contribution to $S$ \cite{Gates:1991uu}) this scenario could be viable, since there are areas of the parameter space that give small enough $S$ but large $T$. The negative contribution to $T$ would then move these points into the experimentally allowed region. In this scenario the Higgs production cross section via gluon fusion would be greatly enhanced, approximately $\sim 25$ times the SM cross section, but we will not explore this case any further.

\subsection{TC(3,2-2S) and Two New Lepton Generations}
\label{twoleptons}

The case where one lepton generation arises in TC(2,2-2S) has been studied in \cite{Antipin:2009ks}. Here we complement that study by considering the case where two lepton generations arise in TC(3,2-2S). For simplicity we consider only the leptons with Dirac mass terms; as discussed and shown explicitly in \cite{Antipin:2009ks}, the corrections due to Majorana mass terms are small. A notable feature of neutrino Majorana mass terms is that they allow for negative contributions to $T$-parameter which is not possible with Dirac mass terms. On the other hand, the present data clearly favors positive contribution to $T$ so this feature is not of immediate interest to us. However, as already noted in connection with new QCD quarks, there may arise cases where additional BSM fields are present and contribute positively to $T$, and then new Majorana neutrinos would be useful in masking such contributions.

In addition to the doublet of techniquarks, we now have the total of four new fermions and thus four different masses. The resulting $(S,T)$-spectrum is shown in the left panel of Figure \ref{twoleptons}. The green points correspond to points where the new generations are hierarchical so that both of the fifth generation leptons are heavier than either one of the fourth generation leptons. The blue crosses correspond with parameter values where at least one of the fourth generation leptons is heavier than at least one of the fifth generation leptons. We see that small values of $S$ and $T$ are accessible if this kind of hierarchy is required or not. The right panel of Figure \ref{twoleptons} shows the allowed mass differences of the fourth and fifth generation when $S$ and $T$ are required to overlap with the ellipsis of Figure \ref{quarks}; again we have checked that $U<0.05$ for all points shown in the plots. Both generations may have masses of a wide range independent of each other, but the maximal allowed mass difference within one generation is affected by the mass difference of the other. This is because the mass difference of each generation contributes independently to the $T$-parameter, so if one generation has a large mass difference and hence large contribution to $T$, the other one is restricted to smaller values of mass difference. The maximal allowed mass difference within either generation is about 120 GeV. We also note that while the mass difference in either one of the generations may have either negative or positive sign, i.e. either the charged lepton or the neutrino is allowed to be heavier, the area where both mass differences are negative is ruled out. This means that at least one of the charged leptons must be heavier than the corresponding neutrino. If stabilized by additional symmetry, such neutrino can provide a dark matter candidate \cite{Kainulainen:2009rb}.

\begin{figure}
\includegraphics[width=0.49\textwidth]{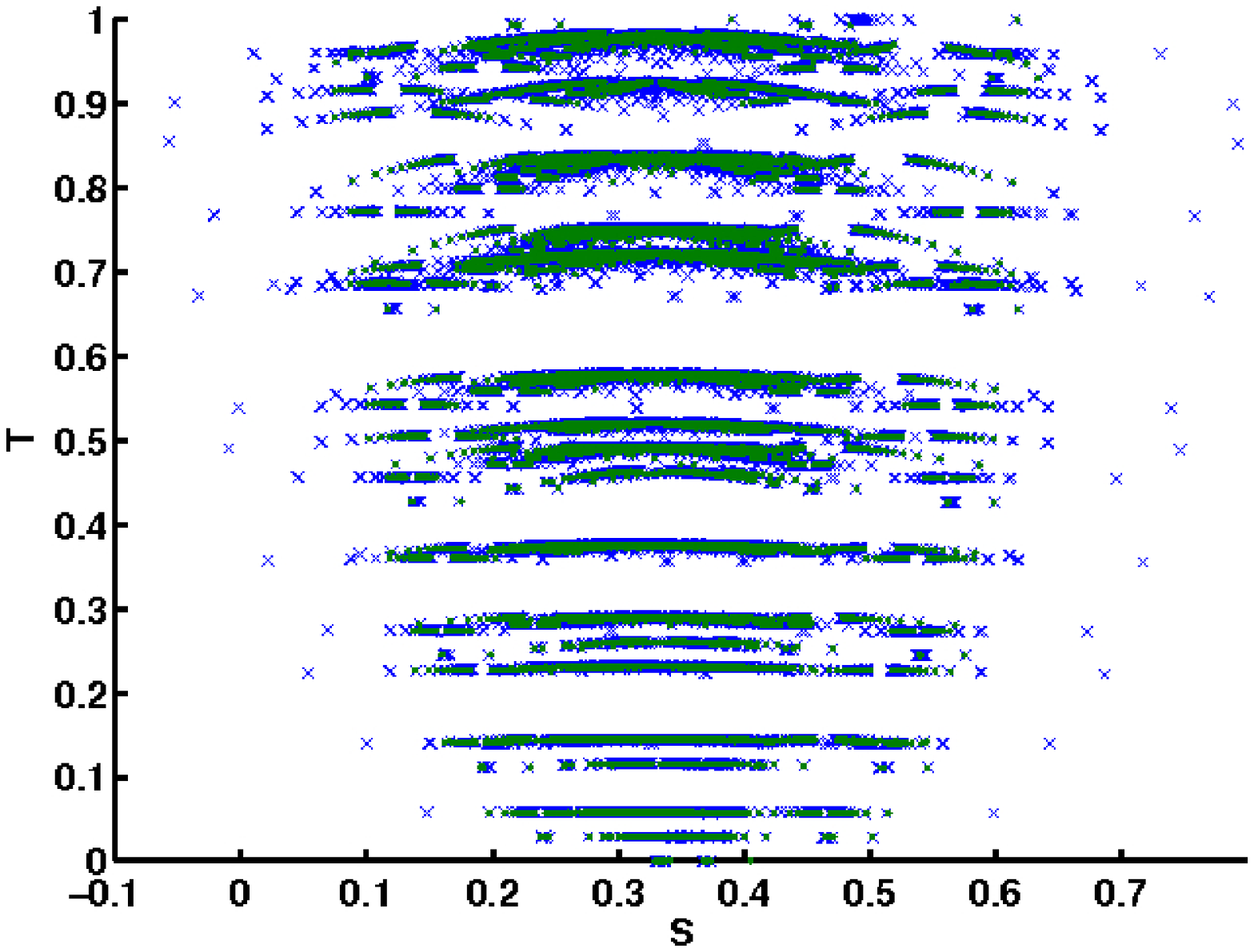}
\includegraphics[width=0.49\textwidth]{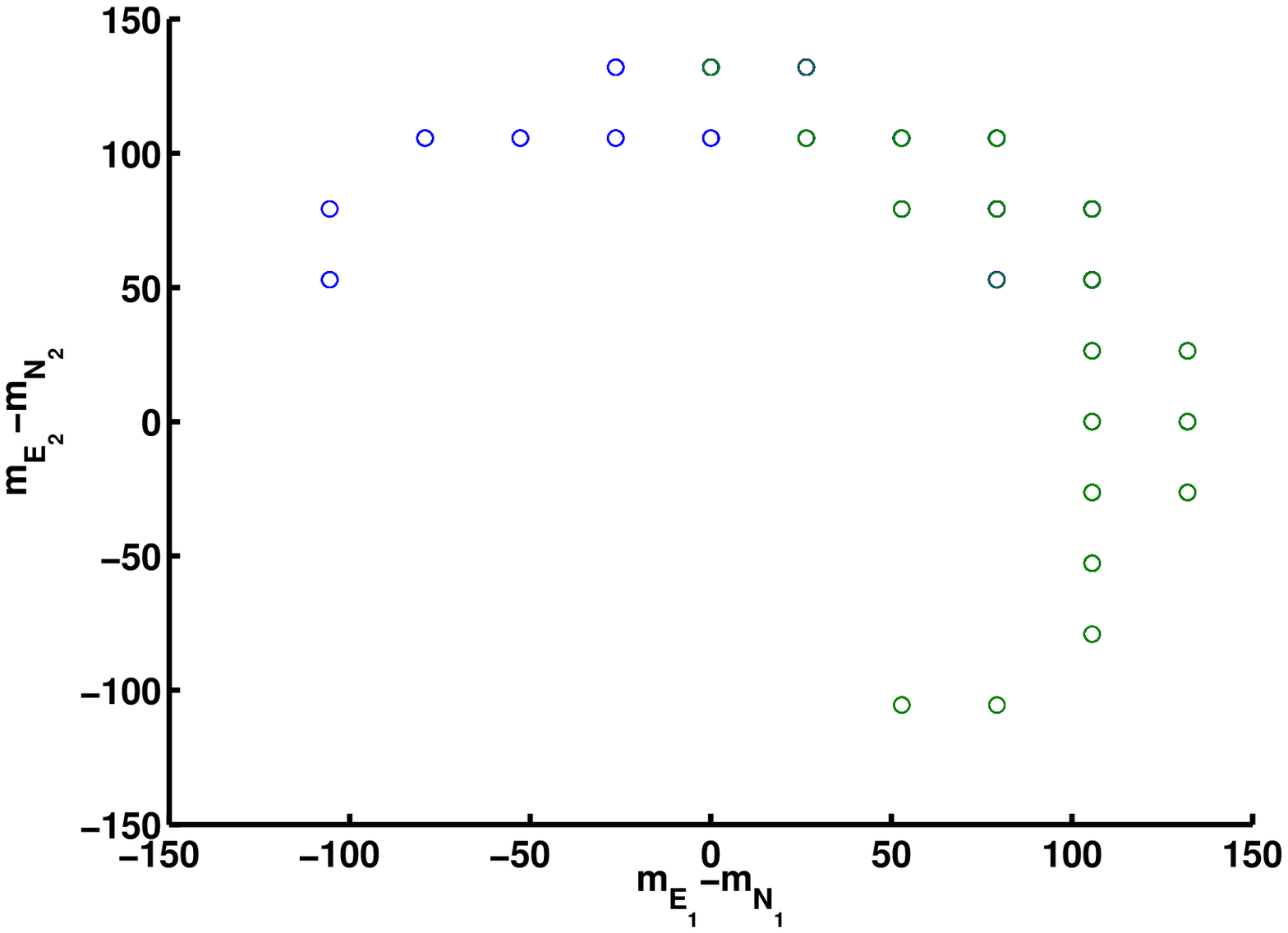}
\caption{Left panel: The $(S,T)$-spectrum of two new SM-like lepton families. The green points correspond to points of the parameter space, where the fifth generation is heavier than the fourth generation, and blue crosses are points where this is not the case. Right panel: The allowed mass differences of the fourth and fifth generation leptons when $S$ and $T$ are required to overlap with the experimental constraint shown in Figure \ref{quarks}.}
\label{twoleptons}
\end{figure}

The collider signatures from two new heavy lepton generations with quantum numbers of an SM-like lepton generation are expected to be similar as the ones considered in \cite{Antipin:2009ks}. Depending on the masses, the fifth generation contribution can be either substantial or negligible. Since there are no new QCD quarks, the dominant channel for Higgs production at LHC will not be enhanced. On the other hand there are now new states available for its decay; especially the neutrinos if light enough can play a significant role as discussed in \cite{Antipin:2009ks}.

Finally, as the constraints from the precision data suggest, there is a possibility where within one of the new lepton generations the charged state is lighter than the massive neutrino. In such a case the charged lepton can only decay via mixing with the lighter generations and depending on the strength of this mixing can lead to interesting signals. We discuss this possibility in more detail in the following subsection.

\subsection{TC(2,2-2S) and Doubly Charged Leptons}
\label{dbchargeleptons}
\begin{figure}
\includegraphics[width=0.49\textwidth]{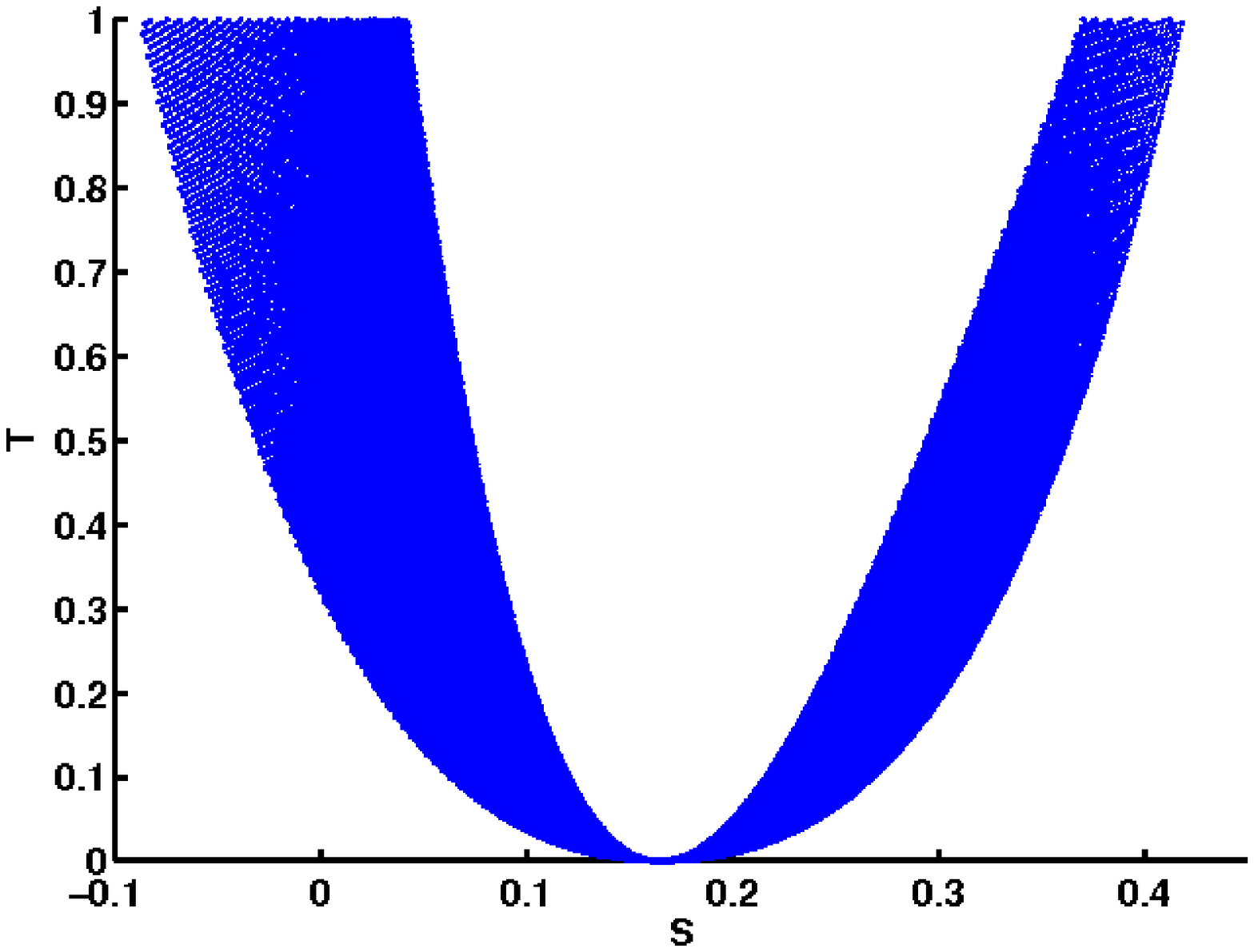}
\includegraphics[width=0.49\textwidth]{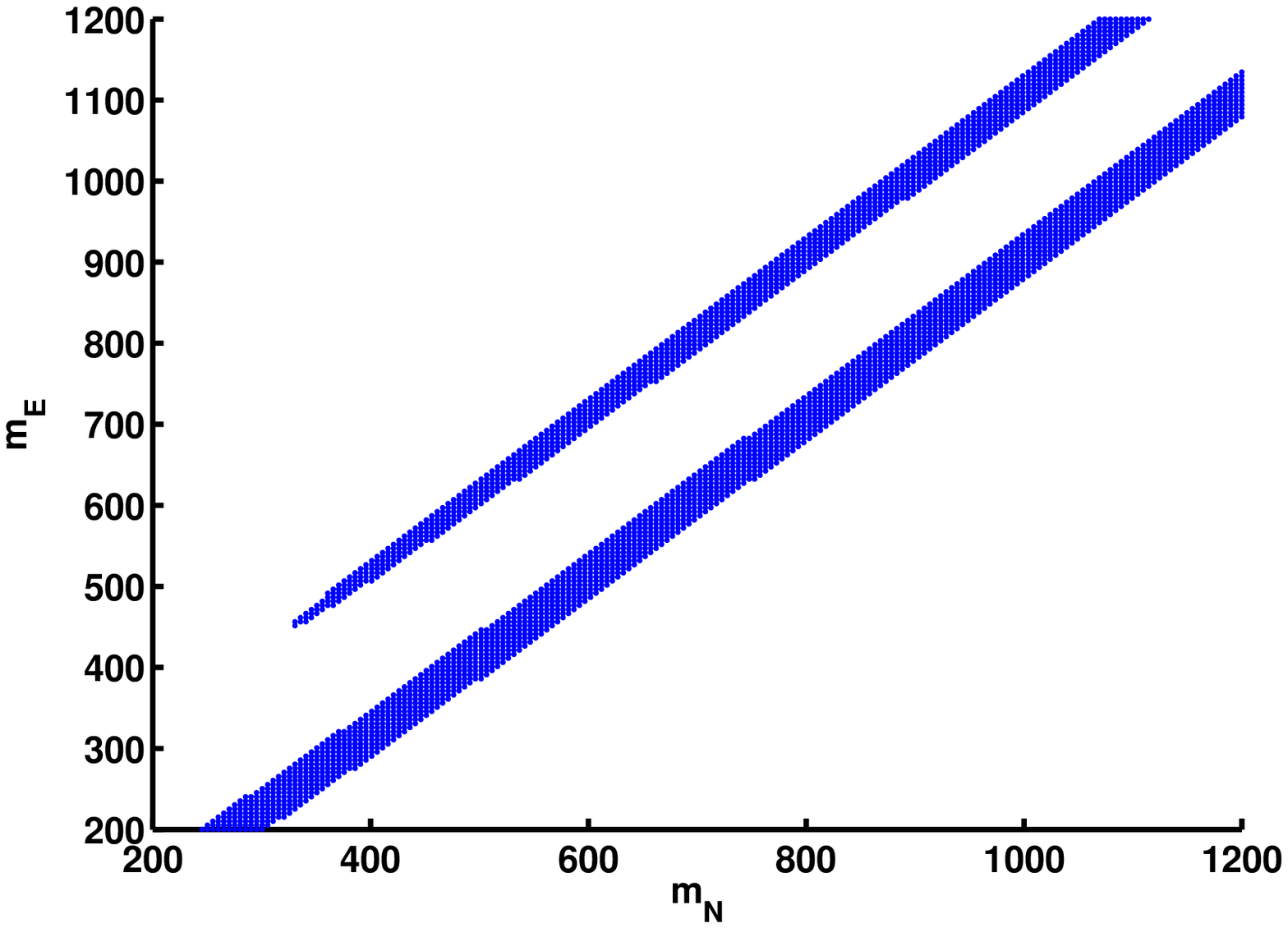}
\caption{Left panel: The $(S,T)$-spectrum of one doubly charged ($Y=3/2$) lepton family. Right panel: The allowed masses of the leptons.}
\label{dbchlept}
\end{figure}

For completeness, we also present the results for the case of one doublet of leptons with charges 2e and e. The $(S,T)$-spectrum of a doubly charged lepton family is shown in the left panel of Figure \ref{dbchlept}. We see that there is again a large overlap with experimentally allowed values. Requiring $S$ and $T$ to lie within the ellipsis of Figure \ref{quarks}, we obtain the mass region shown in the right panel of Figure \ref{dbchlept}. Here $m_N$ is the mass of the new ``neutrino'', which now has a charge of $-1$, and $m_E$ is the mass of the $-2$-charged new ``electron''. Again, the scale of the lepton masses is not limited by the oblique corrections, but the relative mass difference is. As we see from the right panel of Figure \ref{dbchlept}, the precision data favors a small mass splitting, $\Delta m\sim 100$ GeV, within the doublet. Either one of the new leptons may be the heavier one, unless they are both relatively light in which case it is the ''neutrino'' which has to be slightly heavier.

Let us briefly discuss the observational consequences of such exotic lepton multiplet. Being charged, both of these states are easy to observe. Since the precision data allows either for $m_E>m_N$ or $m_E<m_N$ let us consider each ordering in turn. The fourth generation leptons can couple to the ordinary leptons via the following gauge invariant but lepton number violating operator,
\be
{\mathcal{L}}_{\Delta L\neq 0}=y_{42}\overline{L}_L^{(4)}(i\tau_2\Phi^\ast)\mu_R,
\ee
where $L^{(4)}=(N,E)$ is the fourth generation lepton doublet, $\Phi$ is an effective (composite) SM-like Higgs doublet and we assumed mixing with only the second generation. 

First, if we assume that $m_E>m_N$, then the doubly charged state is expected to decay via $E\rightarrow NW^\ast$ and $E\rightarrow \mu W^\ast$. The first one is probably favored over the second one, eventhough there is significant phase space suppression due to the fact that the precision constraints require the masses $m_E$ and $m_N$ to be close to each other. Since $N$ is the lightest member of the doublet, it can only decay via mixing with the ordinary leptons and the resulting phenomenology heavily depends on the magnitude of this mixing. The lifetime of $N$ could vary from $10^{-12}$ s, which would allow $N$ to leave charged track for detection, to absolute stability if the mixing was strictly zero. Having a singly charged heavy stable particle would be disastrous for standard cosmology since it would lead to formation of anomalously heavy hydrogen. For metastable charged particle the cosmological constraints are due to the contribution to the cosmic microwave background from the photon emission associated with the decay of $N$, its effects on the Big Bang Nucleosynthesis  and from its contribution to the dark matter abundance \cite{Sher:1992yr,Holtmann:1998gd}. These analyses are consistent with lifetimes as big as of 10-100 years. This discussion also holds for the scenario of Sec. \ref{twoleptons} when the charged member of either new lepton doublets is lighter than the corresponding massive neutrino.

The other possibility $m_E<m_N$ is equally interesting. Here the beginning of the above discussion applies with $E\leftrightarrow N$: The heavier $N$ will decay to $EW^\ast$ or via mixing to the SM leptons, and then the $E$ can only decay via mixing. However, in this case also the possibility that $E$ is absolutely stable is allowed.  As explained in \cite{Khlopov:2007ic}, after its formation in Big Bang Nucleosynthesis, $^4{\rm{He}}$ screens the doubly charged $E$ into composite ``atoms'' ($^4$He$^{++}E^{--}$) and these primordial neutral objects with normal nuclear interactions may saturate the abundance of dark matter and evade direct observation constraints; for further details we refer to \cite{Khlopov:2007ic}.

\section{Unification}
\label{unification}

Finally, let us discuss how extra matter generations, together with techniquarks, also affect the running of the SM coupling constants at one-loop. Generally, the evolution of the coupling constant $\alpha_n$ of an SU$(n)$ gauge theory at one loop level is controlled by
\begin{equation}\label{running}
{\alpha_{n}^{-1}(\mu) = \alpha_{n}^{-1}(M_Z) - \frac{b_n}{2\pi}\ln
\left(\frac{\mu}{M_Z}\right) \ .}
\end{equation}
For SM we have three coupling constants corresponding to  SU$(3)\times$SU$(2)\times$U$(1)$ for $n=3,2,1$. In the above equation the first coefficient $b_n$ of the beta function is 
\begin{equation}
 b_n = \frac{2}{3}T(R) N_{wf} + \frac{1}{3} T(R')N_{cb} -
 \frac{11}{3}C_2(G) \ .
\label{eq:bn}
\end{equation}
where $T(R)$ is the Casimir of the representation $R$ to which the
$N_{wf}$ Weyl fermions belong and $T(R')$ is the Casimir of the representation $R'$ to which the $N_{cb}$ complex scalar bosons belong.  Finally, $C_2(G)$ is the quadratic Casimir of the adjoint representation of the gauge group.

Now we require the SM coupling constants to unify at some very high energy scale $M_{GUT}$. This means that the three couplings are all equal at the scale $M_{GUT}$, i.e. $\alpha_3(M_{GUT})= \alpha_2(M_{GUT})=\alpha_1(M_{GUT})$ with $\alpha_1= \alpha/(c^2\cos^2 \theta_W)$ and $\alpha_2 = \alpha/ \sin^2\theta_W$, where $c$ is a normalization constant to be determined by the choice of the unifying group (like the paradigmatic SU(5)).
Assuming one-loop unification using Eq.~(\ref{running}) for $n=1,2,3,$ one
finds the following relation 
\begin{equation} 
\label{unification}
B \equiv
\frac{b_3-b_2}{b_2-b_1} = 
      \frac{\alpha_3^{-1} -\alpha^{-1}\sin^2\theta_W}
           {(1+c^2)\alpha^{-1} \sin^2\theta_W-c^2\alpha^{-1}} \ .
\end{equation}
In the above expressions the Weinberg angle $\theta_W$, the electromagnetic coupling constant $\alpha$ and the strong coupling constant $\alpha_3$ are all evaluated at the scale $\mu=M_Z$. For a given particle content, we denote the LHS of Eq.~(\ref{unification}) by $B_{\rm theory}$ and the RHS by $B_{\rm exp}$. Whether $B_{\rm theory}$ and $B_{\rm exp}$ agree is a simple way to check if the coupling constants unify. However, finding a convergence of all coupling constants at a common scale is not enough; to have the proton decay under control the unification scale has to be sufficiently large.
With one-loop running the unification scale is given by the expression
\begin{equation}
M_{\rm GUT} = M_Z \exp 
       \left[ {2\pi \frac{(1+c^2)\alpha^{-1}\sin^2\theta_W-c^2\alpha^{-1}}
                         {b_2-b_1}} \right]\,. 
\label{massu}
\end{equation}

To be specific, we will use the experimental values from ref.~\cite{Amsler:2008zzb}: $\sin^2 \theta_W (M_Z) = 0.23119\pm 0.00016$, $\alpha^{-1}(M_Z) = 127.909 \pm 0.019$, $\alpha_3(M_Z) =0.1217\pm 0.0017$ and $M_Z = 91.1876\pm 0.0021$ GeV. For concreteness we consider SU(5)-type unification and take $c=\sqrt{3/5}$. Also, we
are considering only the unification of the SM couplings and not the technicolor
coupling. For the running of all four gauge couplings we refer to the literature
\cite{Gudnason:2006mk}. With these numerical values we find from Eqs.~(\ref{unification}) and (\ref{massu}) the following conditions for a successful 1-loop unification:
\begin{eqnarray} B_{\rm theory} = 
\frac{b_3-b_2}{b_2-b_1} \approx 0.73 \quad {\rm and }\quad
{M_{\rm{GUT}} \approx M_Z \exp\left[\frac{185}{b_1-b_2}\right] \gsim 10^{15}{\rm GeV}\,. }
\label{eq:bounds}
\end{eqnarray}

The key feature affecting unification is that at one-loop the contributions to $b_3 - b_2$ or $b_2 - b_1$ emerge only from particles {\em not} forming complete representations of the unifying gauge group (like the five and the ten dimensional representations of $SU(5)$)  \cite{Li:2003zh}. For example the gluons, the weak gauge bosons and the Higgs particle of the SM do not form complete representations of $SU(5)$ but ordinary quarks and leptons do.

After these preliminary remarks, let us consider the models we have introduced. Instead of the Higgs sector of the SM we have the techniquarks with different hypercharge assignments and these are connected with the number of QCD quark- or SM lepton-like new matter fields through cancellation of gauge anomalies as explained in Sec. \ref{models}. We then find From Eq.~(\ref{eq:bn}): 
\begin{eqnarray}
b_3 & = & 4+\frac{4}{3}N_D(q) -11 \nonumber\\
b_2 & = & 4+\frac{1}{3} (3N_D(q)+N_D(l)+d_{\rm{TC}}N_D(Q)) - \frac{22}{3} \nonumber\\
b_1 & = & 4+\frac{2}{5}(\frac{11}{6}N_D(q)+\frac{3}{2}N_D(l)+d_{\rm{TC}}(2Y^2(Q_L)+Y^2(U_L)+Y^2(D_L)),
\label{gaugeb}
\end{eqnarray}
where we used $T(R)=1/2$ for the fundamental representation and appropriate $N_{wf}$ for the corresponding gauge groups. Also, for for $U(1)_Y$ gauge group and given matter field, we have $T(R)=c^2Y^2$, where $Y$ is the corresponding hypercharge. Furthermore, $N_D(i)$ denotes the number of new doublets of QCD quarks, SM-like leptons and techniquarks for $i=q,l,Q$, respectively, and $d_Q$ denotes the dimension of the technifermion representation with respect to the technicolor gauge group. For the models we have considered, $N_D(Q)=1$. The first term equal to four in each factor $b_i$, $i=1,2,3$ above corresponds to the contribution from the  three generations of ordinary quarks and leptons. This contribution obviously drops out in the differences $b_i-b_j$, in agreement with the fact that they form complete representations of the unifying gauge group. 

Recall that anomaly cancellation in these models fixes the difference $\Delta N=N_D(l)-N_D(q)$. It therefore turns out useful to eliminate, say, $N_D(l)$, in favor of $\Delta N$ so that  the equations (\ref{gaugeb}) become
\begin{eqnarray}
b_3 & = & 4+ \frac{4}{3}N_D(q) -11 \nonumber\\
b_2 & = & 4+ \frac{1}{3} (4N_D(q)+\Delta N+d_{\rm{TC}}) - \frac{22}{3} \nonumber\\
b_1 & = & 4+ \frac{2}{5}(\frac{20}{6}N_D(q)+\frac{3}{2}\Delta N+d_{\rm{TC}}(2Y^2(Q_L)+Y^2(U_L)+Y^2(D_L)),
\label{b_equations}
\end{eqnarray}

From this form it is furthermore easy to see that $B_{\rm{theory}}$ depends only on the difference $\Delta N$ and not on the numbers $N_D(l)$ and $N_D(q)$ independently. This allows us to study unification for all these scenarios very easily.

\begin{table}[hbt]
\caption{The values of $B_{\rm{theory}}$ for different cases of SM-like BSM matter considered in the text. The two last columns give the corresponding values when further matter fields in the adjoint representation of the SM gauge groups are added (see text for details).}
\label{unification_table}
\begin{center}
\begin{tabular}{|c|c|c|c|c|c|c|}
\hline
$|Y(Q_L)|$ & $d_{\rm{TC}}$ & $\Delta N$ & $M_{\rm{GUT}}$ & $B_{\rm{theory}}$ &$M_{\rm{GUT}}^{\rm{new}}$ &$B_{\rm{theory}}^{\rm{new}}$\\
\hline
1/6 & 3 & 1 &8.2 $\times 10^{12}$ & 0.68   &2.2 $\times 10^{15}$   &0.72 \\
1/6 & 3 & -1 &6.0 $\times 10^{13}$& 0.63   &4.5 $\times 10^{16}$   &0.67 \\
1/6 & 6 & 2 &8.2 $\times 10^{12}$& 0.86    &2.2 $\times 10^{15}$   &0.94\\
1/6 & 6 & -2 &6.0 $\times 10^{14}$& 0.80   &1.8 $\times 10^{18}$   &0.88 \\
\hline
1/2 & 3 & 3 &9.0 $\times 10^{10}$& 0.63    & 3.4 $\times 10^{12}$   &0.66\\
1/2 & 3 & -3 &8.2 $\times 10^{12}$& 0.50   & 2.2 $\times 10^{15}$   &0.50\\
1/2 & 6 & 6 &3.9 $\times 10^{9}$& 0.73     & 4.9 $\times 10^{10}$   &0.76\\
1/2 & 6 & -6 &8.2 $\times 10^{12}$& 0.50   & 2.2 $\times 10^{15}$   &0.50\\
\hline
\end{tabular}
\end{center}
\end{table}

In the SM without any BSM fields one obtains $B_{\rm{theory}}=0.53$ and hence very poor unification. From Table \ref{unification_table} we immediately see that adding more matter with SM-like quantum numbers helps to achieve better unification although most of the cases are still far from benchmark cases like the one provided by supersymmetry. Moreover, the value of the unification scale, $M_{\rm{GUT}}$, is too small and endangers the proton stability. Hence it seems that adding only SM-like matter is not the optimal route to unification here. 
 
Since fermions are natural in our framework, one can entertain the thought of adding a modest number of matter carrying quantum numbers different from the SM matter fields. For example, we can adapt the idea pursued in \cite{Gudnason:2006mk}, i.e. add one Weyl fermion in the adjoint representation of SU(3) of QCD color and one Weyl fermion in the adjoint representation of SU$_L$(2). These particles carry zero hypercharge and hence do not affect the anomaly cancellation considered in Sec. \ref{models}. Their contributions in (\ref{b_equations}) is a positive contribution $+2$ and $+4/3$ to $b_3$ and $b_2$, respectively.

With the adjoint fermions on top of the already considered SM-like matter fields we obtain $M_{\rm{GUT}}^{\rm{new}}$  and $B_{\rm{theory}}^{\rm{new}}$ tabulated in the second-to-last and last column of Table \ref{unification_table}, respectively. The values in the first row correspond to the model studied in \cite{Gudnason:2006mk}. From the tabulated values we observe that with the adjoint fermions the values of $B_{\rm{theory}}$ remain close to the values obtained with only SM-like matter fields, but the value of $M_{\rm{GUT}}$ is enhanced and most of the scenarios are viable when contrasted with the requirement from proton stability.
 
While we do not add much emphasis on the unification properties of these models, it is instructive to see that one can construct extensions of the SM with simple matter content which address several currently favored model building paradigms: First, the models which we have considered are natural in the sense that they contain only fermionic matter and gauge fields. Second, the Technicolor sector or the new leptons, if introduced, may provide suitable dark matter candidates carrying conserved quantum number. Third, some tendency towards one-loop unification of the SM coupling constants can be obtained. Our results seem to indicate that adding only SM-like matter fields on top of the underlying Technicolor model is insufficient for unification and inclusion of further non-standard matter fields is likely needed (like the possibility proposed in \cite{Gudnason:2006mk, Kainulainen:2010pk}).

\section{Conclusions and Outlook}

A simplest extension of the SM is the appearance of a new matter generation. On the other hand, if the electroweak symmetry is broken dynamically by new strong dynamics, new SM-like matter fields can naturally emerge as a result of the saturation of gauge and global anomalies. In this paper we have exhibited several concrete model frameworks where non-sequential next generation arises, shown how the masses within the new generations are constrained by existing data and discussed possible collider signatures and implications on unification of the SM coupling constants.

In particular we have presented models where one or two lepton generations but no new QCD quarks appear. As another simple alternative we studied the case where no new leptons but instead a fourth QCD generation appears. We showed that both of these cases are compatible with present constraints on oblique corrections and outlined their consequences for collider experiments. In particular the case with one QCD generation is appealing due to its implications on flavor physics in connection with the observed CP-B phenomena. The cases with new leptons on the other hand provide a possible simple WIMP candidate as a heavy neutrino. Both cases lead to novel signatures at colliders. In addition to these simple extensions we also briefly considered the cases where exotic doubly charged leptons arise. 

Considering the effects of the non-sequential matter generations on the SU(5)-type unification of the SM coupling constants we found that all of the models we considered led to too small unification scale hence endangering proton stability. This problem was shown to be circumvented via inclusion of few matter fields transforming in the adjoint representation of QCD color SU(3) and weak SU$_L$(2) gauge groups.    

Despite its simplicity, the existence of the next generation of SM-like (or more general) fermionic matter has intriguing consequences if observed at the LHC. First and foremost, it would strengthen the role of naturality as a fundamental principle for Beyond Standard Model physics. Consequently, it would hint towards dynamical origin of the electroweak symmetry breaking. The latter becoming especially tempting expectation if a non-sequential new matter generation was discovered as we have tried to emphasize in this paper.

\section*{Acknowledgements} MH thanks M. Ehrnrooth foundation for financial support.

\end{document}